# Heritable Non Genetic Information That is Independent of DNA and That Governs Organismal Development, Tissue Regeneration, and Tumor Architecture

*David H. Nguyen, PhD; Daniel Gallegos, BS
*Tissue Spatial Geometrics Laboratory, Concord, CA

Corresponding author: DHN, dave@tsg-lab.org

## Abstract

Numerous studies of the tumor microenvironment, interspecies xenografting, and limb regeneration suggest the existence of a Tissue Spatial Code (TSC) that controls tissue structure in a quasi-epigenetic fashion. The term “epigenetic” is an inadequate label for the concept that encompasses the TSC, because this information does not act upon DNA or chromatin via post-translational modifications (i.e. DNA methylation, histone acetylation). A broader term is needed to capture the diversity of three-dimensional (3D) spatial codes in biology. One such term is Heritable Nongenetic Information (HNI, pronounced “honey”), which encompasses the TSC. The term “heritable” is appropriate, because this information is passed onto offspring, otherwise it would have disappeared during evolution. Another reason for the heritability of HNI is that the spatial information observed in tissues is not reducible to the laws of physics, meaning structures like epithelial tubes or neural circuits do not spontaneously form in an aqueous solution; pre-existing physiological information in the microenvironment is necessary. In summary, HNI is defined as: heritable, instructional information that affects molecular-, cellular-, tissue-, or organ-level function without having to post-translationally modify DNA or chromatin in order to be inherited.

## Outline

## A. What is Heritable Nongenetic Information?

**How is Heritable Nongenetic Information Different Than Epigenetic Information?**
Numerous studies of the tumor microenvironment, interspecies xenografting, and limb regeneration suggest the existence of a Tissue Spatial Code (TSC) that controls tissue structure in a quasi-epigenetic fashion. The term “epigenetic” is an inadequate label for the concept that encompasses the TSC, because this information does not act upon DNA or chromatin via post-translational modifications (i.e. DNA methylation, histone acetylation). A broader term is needed to capture the diversity of three-dimensional (3D) spatial codes in biology. One such term is Heritable Nongenetic Information (HNI, pronounced “honey”), which encompasses the TSC. The term “heritable” is appropriate, because this information is passed onto offspring, otherwise it would have disappeared during evolution. Another reason for the heritability of HNI is that the spatial information observed in tissues is not reducible to the laws of physics, meaning structures like epithelial tubes or neural circuits do not spontaneously form in an aqueous solution; pre-existing physiological information in the microenvironment is necessary. In summary, HNI is defined as: heritable, instructional information that affects molecular-, cellular-, tissue-, or organ-level function without having to post-translationally modify DNA or chromatin in order to be inherited.

**How is Heritable Nongenetic Information Distinct from “Nongenetic Inheritance”?**
It is worth noting that other scientists are attempting to unify the complexity of non-genetic inheritance, but that their concept is distinct from HNI. Adrian-Kalchhauser et al. (2020) described non-genetic inheritance (NGI) in the form of epigenetic modifications to DNA and chromatin, using the term Inherited Gene Regulation (IGR) as the unifying concept for nongenetic inheritance in evolution and ecology. However, HNI is defined as not including epigenetics, because the various types of HNI discussed in this article (the tissue spatial code, the molecular environment spatial code, species specific DNA autonomy) do not directly modify DNA nor the proteins that comprise chromatin. As pointed out by Edelaar et al. (2021), the concept of IGR is relevant but too DNA-centric to fully encompass the various mechanisms of nongenetic inheritance in ecology. Edelaar provided several examples of inheritance mechanisms that do not involve DNA, including learned behaviors by offspring who watch their parents, the transfer of parental microbiomes to offspring, prions that are transferred from parent to offspring, and preformed cellular structural templates that operate during cell division irrespective of DNA. We are pleased to add HNI — and under it: TSC and the molecular environmental spatial code — to this ongoing discussion of nongenetic information, especially from the perspective of spatial information in tissue structure.

HNI is also distinct from the concept of nongenetic information described by Bonduriansky et al. (2011), which describes transmission of information from parental behavior and environmental stimuli. Figure 1 presents a classification of nongenetic information across HNI,

Adrian-Kalchhauser's (2020) NGI, and the nongenetic inheritance described by Bonduriansky (2011) and Edelaar (2021).

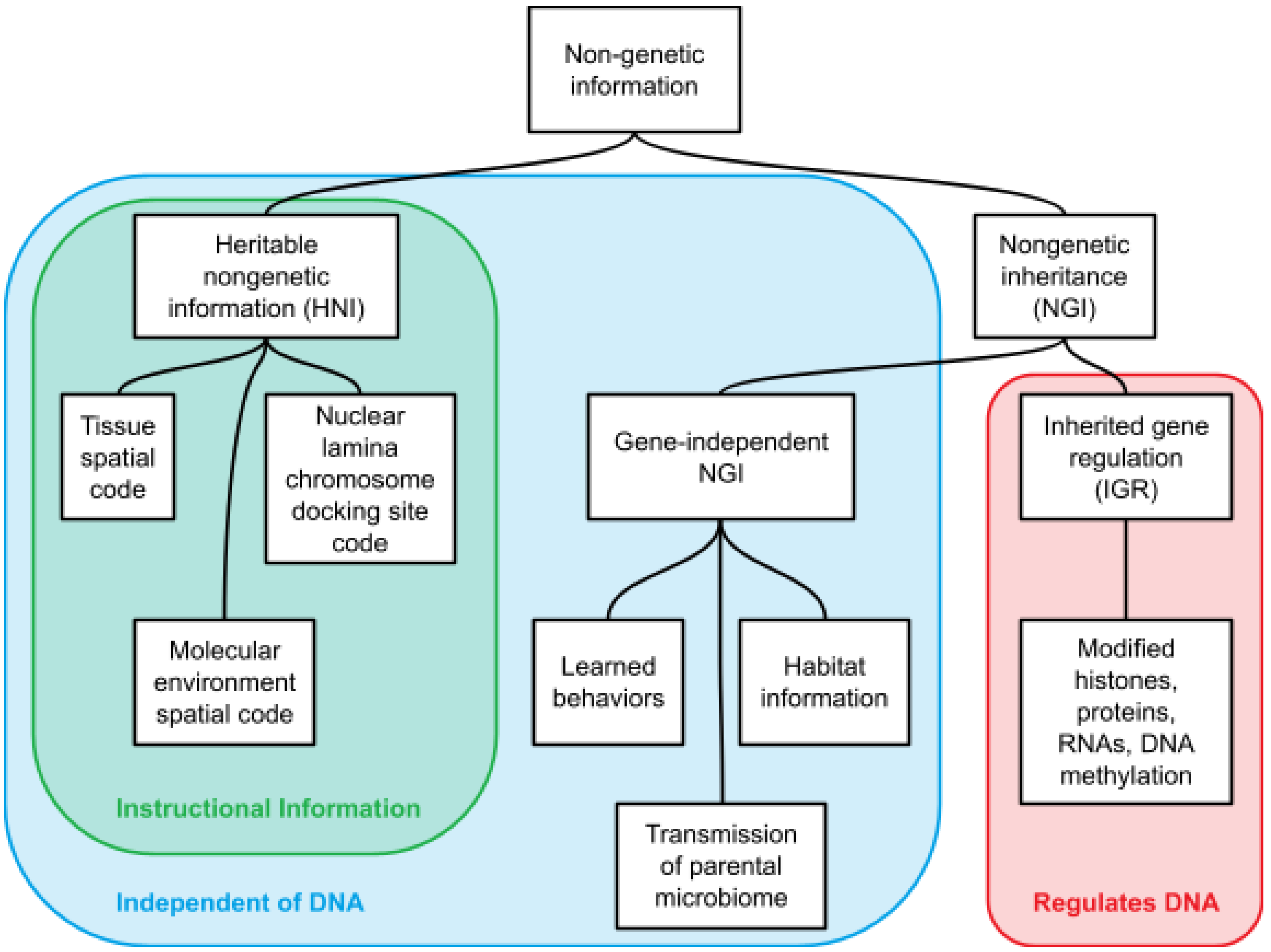

**Figure 1.** A hierarchy of nongenetic information. Several examples of each type of nongenetic information are given. These examples are not exhaustive. HNI is defined in this manuscript. Epigenetic NGI is defined in Adrian-Kalchhauser et al. (2020). Gene-independent NGI is described in Edelaar et al. (2021) and Bonduriansky et al. (2011).

**Table 1.** Comparison Between Heritable Nongenetic Information and Nongenetic Inheritance.

| | **Type of non-genetic information** | |
|---|---|---|
| **Feature** | **HNI** | **NGI** |
| Information carriers | TSC, molecular environment spatial code, nuclear lamina chromosome docking site code | Modified histones, proteins, RNAs, DNA methylation (Adrian-Kalchhauser, 2020)<br><br>Prions, learned behaviors, parental microbiome, habitat information (Edelaar, 2021; Bonduriansky, 2011) |
| Nature of information | Instructs the assembly and operation of molecular, cellular, tissue, or organ systems | Regulates DNA or alters behavior or biomolecule function |
| Transgenerational persistence | Transmitted with high fidelity across numerous generations | Persistent from one to several generations (Adrian-Kalchhauser, 2020; Edelaar, 2021; Bonduriansky, 2011) |
| Relation to DNA | Independent of DNA | Either independent of DNA or an epigenetic regulator of DNA |

**The Benefits of Quantifying 2D and 3D Spatial Codes**
The first step in cracking a code is to determine the syntax and grammatical rules of that code. For DNA, Erwin Chargoff discovered that the ratio of adenine to thymine or guanine to cytosine was similar across different species. Eventually, Watson & Crick discovered the double helical structure of DNA. Further work then revealed that this double-helical structure harbored instructions for genes to carry instructions from making proteins, and that specific sequences of DNA base pairs encoded information for promoter regions, transcriptional bindings sites, splicing sites, etc.

A tissue spatial code, however, is not semantic information, such as that of DNA sequences or the English language, and thus needs specific tools for quantifying regularities in spatial grammar. Recurring quantitative patterns in tissue regeneration or neoplasia will reveal some form of a grammar. A system that can describe encoded 3D spatial information for one subfield of biology will be adaptable to other subfields of biology, and science in general, that deal with the same challenges. Work to develop such a system is ongoing at the Tissue Spatial Geometrics Lab, which has developed five algorithms for quantifying 2D spatial information in tissues and tumors (Nguyen, 2017a; 2017b; 2017c; 2017d), one of which can uniquely quantify 3D shapes in the form of consecutive 2D slices (Nguyen, 2018).

There are multiple benefits to developing an objective, quantitative framework for describing how 2D and 3D information is encoded and how this correlates with biological function.

1. Help crack the syntax and grammar of a code. Codes have recurring units of structured information, similar to words in a language, which are referred to as syntax. These units of information interact with each other according to further rules, similar to the grammatical rules of a language. Recurring units of information, whether in a specific sequence of DNA or the shape and arrangement of cells are hints at syntax and grammar of a TSC. The more detailed the understanding of the syntax and grammar of a code, the more information can be learned about the themes, functions, and nuances of the code, even if the code is not fully elucidated.
2. Help trace the flow of information to reveal specific functions of the code. Codes store information, allow the information to be accessed, and allow the information to be translated into specific functions. In other words, codes allow information to not just exist, but to flow for various purposes, such as homeostatic feedback and error correction.
3. The themes underlying syntax, grammar, and flow of 3D codes will reveal insights into evolutionary relationships, microbial interactions, disease progression, and potential therapeutic strategies. Complex systems consisting of a large library of items, that also exhibit recurring rules of interactions between these items, harbor mathematical themes revealing preferences that naturally emerge because of the rules. Human languages are an example of such systems. Quantitative approaches in the study of human languages based on the frequency of words used have revealed insights into how languages evolve and why (Calude, 2011; Pagel, 2007; Atkinson, 2008). The information in DNA is another example of such a system. The Central Dogma of Molecular Biology was a groundbreaking representation of the flow of information from DNA to RNA to protein. However, as scientists continued to elucidate DNA sequence motifs, it was discovered that both coding and non-coding regions were important for cellular function, and that RNA has diverse roles beyond being just messenger RNA. Furthermore, one gene can encode for a polypeptide that can be spliced into several proteins that have distinct functions, so one gene does not always encode for one protein. The ability to trace the flow of information has revealed that the Central Dogma is grossly outdated and needs major revision.

## B. Evidence of a Tissue Spatial Code as a Form of Heritable Nongenetic Information

There is ample evidence that the TSC exists and is independent of the information in DNA. Note that "independent of the information of DNA" does not mean that DNA is unnecessary, because DNA mutation experiments prove that DNA is quite essential for life. The point is that distinct code systems can operate in parallel and be integrated with each other via dynamic feedback loops to achieve the same goals. In principle, there needs to be a minimum set of genes to produce the functional protein machinery that sustains the functions of cells. However, at the tissue level, the TSC can compensate for the inadequacy of heavily damaged DNA within the cells that comprise the tissue.

**Limb Regeneration in Salamanders**

The fact that amputated limbs regrow the same arm and hand structure without a pre-existing embryonic patterning environment is evidence of a tissue spatial code (reviewed in Simon & Tanaka, 2013; McCusker & Gardiner, 2013). Amputating an arm of the salamander results in a new arm that regrows in its place. This extra arm has the same macroscopic features as an original arm. After limb amputation, the movement of existing nerves towards the edge of the injury activates a tissue regeneration program (Endo, 2004). This limb regeneration program is activatable even without the movement of existing nerves, but via exogenous treatment with retinoic acid and growth factors (Vieira, 2019), which suggests that an inherent tissue development blueprint exists in the cells of the wound site. While DNA carries instructions for creating protein machinery, something else -- heritable nongenetic information -- is dictating tissue patterning outside of the embryonic environment.

**Regenerative Abilities of Flatworms**

Flatworms (Planarians) contain stem cells throughout their body, which allows for a small fragment of a worm to regenerate into a new worm, normally functioning worm. Cells within flatworms communicate by sending electrical signals through gap junctions that connect neighboring cells to each other. Temporarily disrupting electrical signals, by treating worms with a chemical that inhibits gap junctions, permanently changes the anatomical structure of worms to be two-headed worms instead of single-headed worms (Durant, 2017). This effect persisted indefinitely even though the genomes of the two-headed worms were identical to that of single-headed worms, because they were both clones of the original worm. This is very compelling evidence that a bioelectrical signal can control anatomical morphology without altering DNA sequences. Planarians have a physiological memory that controls anatomy independently of DNA sequence.

The relatively simple anatomical structure of flatworms makes it an ideal experimental model by which to measure the spatial information of tissue structure, in the form of cellular shape and arrangement. The morphology of neurons has evolved to expedite the sending and receiving of electrical signals. Why wouldn't we expect that epithelial and mesenchymal cells that transmit electrical signals between each other won't be shaped or arranged in a way that optimizes the transmission of these signals? If this spatial optimization is true, it can be measured histologically with the right computational tools.

**Human-Mouse Xenografting Overrides DNA Integrity or Sequence Divergence**
Injecting human B16 melanoma cells into a mouse blastocyst can result in a viable adult mouse that is a mosaic of human and mouse cells throughout its body (Mintz, 1978). This is the first example here of "species specific DNA autonomy" (SSDA). Not only are the melanoma cells non-murine, their DNA harbors many mutations and chromosomal aberrations. Another example of a tissue spatial code dominating orthotopic DNA instructions is the humanized mouse mammary glands. There is a method to inject human mammary epithelial cells along with human mammary fibroblasts into the mammary fat pad of mice. This results in functional human mammary ducts in an immunocompromised mouse (Kuperwasser, 2004). This is another evidence of SSDA, meaning the extracellular/tissue environment can override the DNA of the xenografted cells.

**In Vitro Microenvironmental Programming of Cancer Cells**
Artificially derived extracellular matrix (matrigel) can revert the cancerous phenotype of very aggressive cancer cell lines, a mechanism mediated by Integrin β1 (Weaver, 1997). Breast cancer cell lines that have many mutations and chromosomal aberrations can be coaxed into forming organized acini in culture, just like how premalignant breast cell lines behave. There is also evidence that malignant breast cancer cell lines only respond to growth inhibiting molecules in a 3D environment and not in 2D cell culture despite harboring the same mutations in both contexts (Wang, 1998; Liu, 2004). This is evidence that the extracellular environment can override the instructions in damaged DNA codes.

**Tumor Histology Exhibits Recurring Patterns Identifiable by Pathologists**

One of the oldest lines of evidence that a TSC exists comes from the medical field of tumor pathology. While tumors exhibit a tissue architecture that seems haphazard and random compared to the tissues of their origin, the disorderly architecture does present in the form of recurring patterns. These recurring patterns of tumor architecture make it possible for a pathologist to recognize the common histological subtypes of each solid cancer. Tumors of epithelial origin, such as breast cancer (reviewed Makki, 2015) and lung cancer (reviewed in Kuhn, 2008), often exhibit features that are reminiscent of normal epithelial ducts, such as finger-like tubes referred to as a “papillary” histology when viewed in a two-dimensional slice of the tumor.

## C. Molecular and Subcellular Evidence of Heritable Nongenetic Information

**Natively Unfolded Proteins (NUPs) as Evidence of Heritable Nongenetic Information**
Natively unfolded proteins and protein domains that are natively unfolded, do not exhibit a regular structure until they interact with an appropriate environment (reviewed in Nishikawa, 2009). For example, a NUP may not fold into a regular 3D structure until it interacts with the hydrophobic environment of a cellular membrane. NUPs have also been reported to take on different 3D structures depending on which protein partner they bind to (reviewed in Tompa, 2005). The fact that a polypeptide chain needs spatial information that is independent of the polypeptide's own sequence suggests that the full set of instructions for protein folding -- and thus, function -- is *not* contained in DNA. This is evidence of a "molecular environment spatial code."

**Specific Docking Sites for Each Chromosome on the Nuclear Lamina**
Photobleaching experiments have shown that mammalian chromosomes have specific "addresses" on the inner nuclear membrane (Boyle, 2001; Essers, 2005). The lamina-associated domains (LADs) in DNA contain methylated adenines, and are responsible for chromosome regions that physically locate at the inner nuclear membrane, specifically at the nuclear lamina (reviewed in van Steensel & Belmont, 2017). LADs are ubiquitously distributed across all mammalian chromosomes, yet each chromosome has a preference for specific locations on the nuclear lamina; there must be a code in operation.

## D. Conclusion

The examples of HNI presented in this article underscore the depth of biology that is yet to be revealed. While genetic modification, such as CRISPR, can have off-target effects in medical applications that result in negative side effects, HNI holds the promise of modifying cellular behavior without directly modifying DNA. As with previous discoveries about fundamental mechanisms of how biomolecules, cells, and tissues work, HNI can transform our understanding of life and advance our capabilities in medicine. The challenge of cracking 3D spatial codes in biology will require out-of-the-box thinking and interdisciplinary approaches.